\documentclass[twoside]{article}
\usepackage{fleqn,espcrc2}

\def\beq{\begin{eqnarray}}   \def\eeq{\end{eqnarray}}

\def\centeron#1#2{{\setbox0=\hbox{#1}\setbox1=\hbox{#2}\ifdim
\wd1>\wd0\kern.5\wd1\kern-.5\wd0\fi
\copy0\kern-.5\wd0\kern-.5\wd1\copy1\ifdim\wd0>\wd1
\kern.5\wd0\kern-.5\wd1\fi}}
\def\ltap{\;\centeron{\raise.35ex\hbox{$<$}}{\lower.65ex\hbox{$\sim$}}\;}

\def\gtap{\;\centeron{\raise.35ex\hbox{$>$}}{\lower.65ex\hbox{$\sim$}}\;}

\def\dslash{\not{\hbox{\kern-2pt $\partial$}}}
\def\Dslash{\not{\hbox{\kern-4pt $D$}}}
\def\Oslash{\not{\hbox{\kern-4pt $O$}}}
\def\Qslash{\not{\hbox{\kern-4pt $Q$}}}
\def\pslash{\not{\hbox{\kern-2.3pt $p$}}}
\def\kslash{\not{\hbox{\kern-2.3pt $k$}}}
\def\qslash{\not{\hbox{\kern-2.3pt $q$}}}
\def\epsilonslash{\not{\hbox{\kern-2.3pt $\epsilon$}}}

\newcommand{\newc}{\newcommand}
\newc{\qbar}{{\overline q}}
\newc{\Kahler}{K\"ahler }
\newc{\deltaGS}{\delta_{\rm GS}}
\begin{document}
\sloppy

\begin{flushright}
{\large hep-th/0107263 \\ SCIPP-01/21\\}

\end{flushright}

\title{
String Theory, Large Dimensions and Supersymmetry\footnote{Talk
presented at 30 Years of Supersymmetry \\ Minneapolis, October,
2000}
}
\author{ Michael Dine  \address{Santa Cruz Institute for Particle
Physics\\ University of California (Santa Cruz)\\
Santa Cruz CA 95064  }}

\begin{flushright}
{\large hep-th/0107263 \\ SCIPP-01/21\\}

\end{flushright}
\begin{abstract}
[Invited talk presented at 30 Years of Supersymmetry, Minneapolis,
October 2000]

\noindent
With our current level of understanding, the problem of making
string theory predictions is not one of ``solving" the theory, but
rather of trying to determine whether there are any generic
expectations.  Within this context, we discuss what it would mean
to predict low energy supersymmetry, and consider questions like:
what is the form of low energy CP violation, is unification a
string prediction, and others.

\end{abstract}

\maketitle

\section{SUPERSYMMETRY AND STRING THEORY}

The study of supersymmetry had its beginnings in string
theory, and for much of their history the two subjects have been
strongly linked.
The past five years have witnessed great advances in our
understanding of string theory.  Almost all of these have been
closely tied to supersymmetry.
These include:
\begin{itemize}
\item Dualities among string theories:  Essentially all of the evidence
for
various dualities is based on relations which hold because of
supersymmetry.
\item Non-perturbative formulations of the theory in various limits:
Matrix models\cite{taylorreview}  and
AdS/CFT duality\cite{adsreview}, are based on general statements, but
again virtually all
of the evidence
provided in support of these formulations
can be understood in terms of supersymmetry.  It is not
clear that the Matrix model proposal makes sense without
supersymmetry.
\item  Many other developments, such as the discovery of string
theories without gravity, rely heavily on supersymmetry.
\end{itemize}

Yet at the same time, we don't know whether ``low energy
supersymmetry" is an outcome of string theory.  Indeed, while it
is often claimed that low energy supersymmetry is one of the few
predictions of string theory, we are far from being able to make
such a statement.  This is clear from Dimopoulos's
talk at this meeting.  Large or warped dimensions have been
widely explored over the last two years as an alternative solution of
the
hierarchy problem\cite{earlylargedimensions,largedimensions,rs}.
Most scenarios of this kind do not invoke low
energy supersymmetry in any conventional sense.
Yet the large dimension story itself fits naturally into string
theory (indeed it is hard to make sense of it without string theory).

Over the next decade, low energy supersymmetry will be confirmed or
ruled
out.  If not observed at the LHC, our ideas about supersymmetry and the
hierarchy are simply wrong.   Large dimensions as a solution to the
hierarchy problem are perhaps harder to rule out.  We do not really have

detailed models which permit a precise phenomenology, as we do for
supersymmetry.  Those who
contemplate two large dimensions, for example, are already
confronted with a fundamental scale of order $50$ TeV. Even
with more dimensions, the scales are uncomfortably large from the
perspective of hierarchy.  So, much as for technicolor,  if we accept
this
type of
explanation, it is hard to make
a precise statement of our expectations; we will simply have to
await an experimental discovery (as for technicolor, we can give
some qualitative experimental expectations:  dense sets of states, in
particular, giving rise to surprising processes at high energies).

It would be important to be able to
say:  supersymmetry is a prediction (or not!)
of string theory, within the next few years.
It would be desirable to make a similar statement about large
dimensions.
We might hope to arrive at this point by ``solving" string theory,
and discovering that the
solution has the standard model, low energy supersymmetry,
unification, etc.  But this is unlikely to happen soon.
A more realistic hope is that we might make generic predictions,
such as low energy
supersymmetry, unification, some statements about CP or flavor.  Even
this is a tall order,
given our current level of understanding.

In this talk, I will outline some ways in which we might
be able to make a decisive statement about supersymmetry
in string theory.  I will argue
that there is some evidence that string theory prefers supersymmetry.
The evidence, currently, is tenuous, but there are ways in which
we might attempt to make a firm statement.
It is
less clear how we might make an analogous statement about large
dimensions (if we fail to make a case for supersymmetry).  Short
of providing some sort of solution to the theory, we would need to
provide some compelling argument why such states are preferred.
It seems more likely that the large dimension scenario will remain
a scenario; if the large dimension story is correct,
it will not be predicted, but rather (hopefully) discovered.

While I will raise a number of questions, I will provide
answers only in a few cases, and even these will be, at best,
conjectures.  But I hope to suggest that many questions involved in
relating
string
theory to nature are not totally out of reach.  Among the
questions which I will try to formulate:
\begin{itemize}
\item What is String Theory?  This is a question we understand, in
part, for theories with a large amount of supersymmetry, but for
theories with little or no supersymmetry, this is far less clear.
\item Some good string ground states:  It is often said that the
problem of determining the ground state of string theory is a
dynamical one.  But states with enough supersymmetry are almost
certainly good ground states.
\item Approximate Moduli Spaces and N=1 Supersymmetry: If we are
to formulate the question: does string theory predict low energy
supersymmetry? in a generic way, we must first decide what it would
mean to make such a prediction.
\item  Is Anything the Matter with N=0 Supersymmetry?  Having
characterized the distinction between theories with low energy
supersymmetry and those without, we argue that there is some
evidence that theories without supersymmetry generically suffer
from diseases.
\item  Is Small $\sin (2\beta)$ a Supersymmetry Prediction?  I
will argue that ideas about supersymmetry breaking suggest that
the asymmetry in $B$ decays should either be very small, or quite
close to the standard model value [Recent results from the $B$ factories

have ruled out the small asymmetry hypothesis.]
\item  Some Anthropic Issues:  Given that string theory has many
ground states which are drastically different from what we
observe, it seems quite possible that anthropic considerations
will play some role in determining the ground state in which we
find ourselves.
\item  Coupling Unification:  One of the problems with our
improved understanding of string theory and its strongly coupled
limits, is that it is no longer clear that coupling unification is
a robust prediction.
\item   The Brane World in String Theory:  The brane world
picture is an exciting possibility, which could manifest itself in
experiments at colliders.  Theoretically, it poses many
challenges, some of which are reviewed here.   In the case of
anomaly mediation, we argue that the brane world ideas are not
robust.
\item  The role of holography:  Much of the language I will use
assumes a conventional effective field theory
description at low energies.  As Banks has stressed, more radical
proposals
for the role of supersymmetry in string theory might well yield
different answers to many of these questions\cite{littlelambda}.
He suggests, based on considerations of string theory in De Sitter
space, that the number of degrees of freedom might be a parameter,
which in turn determines the cosmological constant and the degree
of supersymmetry breaking.  He also argues that supersymmetry may
be crucial, at least in some asymptotic sense, to attaining a
consistent picture.
I will not explore this intriguing possibility here (except for some
later comments in the context of vacuum selection and cosmology),
but just acknowledge
that, given our poor understanding of the question of the
cosmological constant, such radical reformulations of our thinking
may well be necessary, and may lead to surprises.
\end{itemize}

\section{What is String Theory?}

Susskind\cite{susskind} has proposed a provocative definition of
string theory:  it is that theory which lives on the moduli space
of supergravity theories with a sufficiently large number of
supersymmetries that we can give a non-perturbative definition
(in some region of the moduli space).
With our current level of understanding, this
means that the number of supersymmetries must be, say, $16$ or $32$, so
that we can give Matrix Model or ADS/CFT definitions.  This view
is provocative precisely because it excludes the world we observe.
I would argue that we can do much better than this.

\subsection{Good String Ground States}

But it is first worthwhile to note that string states in $4-10$
dimensions with eight or more supersymmetries almost certainly
exist, i.e. they are perfectly good ground states of string
theory.  This is simply because it is not possible to write
potentials for the moduli consistent with supersymmetry.
So the problem of understanding why we live in four
dimensions, or why in four d we don't see $N>1$ is not simply a
problem of deciding what is the lowest energy state, or some
other well-posed dynamical question.  We will
engage in some speculations later.

\subsection{A Response to Susskind's Challenge}

We have at least good circumstantial
evidence that in string theory,
states with $N=1$ supersymmetry in four dimensions exist.
More precisely, there exist approximate moduli spaces
with $N=1$ supersymmetry.
This evidence is based on considerations of dualities.
It is much in the spirit of Witten's original discussion
of duality\cite{wittenusc}.

What is meant by an approximate moduli space?  Consider,
e.g., some compactification of the weakly coupled
heterotic string in which
gluino condensation gives rise to a superpotential
for the dilaton,
\beq
W= c e^{-S/b_o}.
\eeq
Here $S = {V \over \ell_s^6 g_s^2}$, where $V$ is the volume of the
compactified space, $\ell_s$ is the string length scale, and $g_s$ is
the
string coupling constant.
In this case, there is a potential for $S$ which goes to zero as
$S \rightarrow \infty$.  Supersymmetry is restored in the limit as
$S \rightarrow \infty$, and in particular, there is one light
gravitino in this limit.

We also understand a great deal about the strongly coupled limit.
The $E_8 \times E_8$
heterotic string goes over to an eleven  theory with
two boundaries\cite{horavawitten}, separated by a distance $R_{11}$.
The characteristic scale of this theory is the eleven dimensional
Planck scale, $\ell_{11}$.
The basic relation between the two pictures is:
\beq
g_s^2 =  {R_{11}^3 \over \ell_{11}^3} ~~~~~   \ell_s^2 = {\ell_{11}^3
\over R_{11} }.
\eeq
In the strongly-coupled picture, we can compactify six of the remaining
dimensions
on, say, a Calabi-Yau space characterized by a radius $R$.  The
moduli now are $R_{11}$ and $R$, or $g_s$ and $R$.  It is
conventional to write these as $S$ and $T$.  At weak coupling,
\beq
S = g^{-2}{V \over \ell_s^6}   ~~~~~  T = {R^2 \over \ell_s^2}.
\eeq
The weak coupling description is valid when
\beq
g_s^{-2} \gg 1 ~~~~~~  {R\over \ell_s} \gg 1.
\eeq
or
\beq
{S \over T^3} \gg 1 ~~~~~~  T \gg 1.
\eeq
At strong coupling:
\beq
T = V^{1/3} {R_{11} \over \ell_{11}^3}     ~~~~~  S = { V \over
\ell_{11}^6}.
\eeq
This description should be valid when
\beq
R_{11} \gg \ell_{11}  ~~~~~~~  V \gg \ell_{11}^6,
\eeq
or ${T^3 \over S} \gg 1 ~~~~~~ S \gg 1$.

It is important that the regions of validity of weak and strong
coupling do not overlap.
However, in the effective low
energy theory, one can calculate, in both regimes, certain
holomorphic quantities (the superpotential and the gauge coupling
function):
\beq
f(S,T) = S + a T + {\cal O}( e^{-T},e^{-S}) \\
W(S,T)= e^{-{(S+aT)\over b}} + \dots
\eeq
One {\it can} pass from weak to strong coupling keeping $S$, $T$ large.
So these quantities {\it should} agree in these two regimes, and
they do\cite{bdhv}!
These checks are non-trivial and arguably at least as
impressive as the agreement in cases with exact moduli spaces and
more supersymmetry.

\section{What is Distinctive About Low Energy Supersymmetry?}

Even without Susskind's challenge, we might have
asked:  in what sense might string theory predict low energy
supersymmetry.  After all, what we ``want" is to argue that
nature is approximately supersymmetric.  But in what sense
would this be a generic outcome of string theory?  What
distinguishes a vacuum with approximate supersymmetry from one
with no supersymmetry?  Typically one speaks of theories in which
``supersymmetry is unbroken at tree level."  But it is unlikely
that the regime of string theory which describes our world is
weakly coupled, so it is unclear why ``tree level" statements
should have any relevance.  The existence of a hierarchy does suggest,
however, that we might sit at a point in an approximate moduli space.
The
approximate moduli spaces which admit low energy supersymmetry are
distinguished by the
fact that, in some limit, they become exactly supersymmetric with
only four supersymmetries.  This contrasts with states
which become supersymmetric, if at all, only in limits with
infinite numbers of supersymmetries (infinite numbers of
gravitinos, for example, in some large radius limit).
This discussion suggests what it would mean to show
that string theory predicts low energy supersymmetry:
one would want to argue that the ground state which describes
the world we observe sits on one of these approximate moduli
spaces.

We have at least given evidence here that such approximate
moduli spaces exist.
The facts suggest that if string theory describes nature,
the vacuum which describes our world sits on such an approximate
moduli space.  How this might come about is the subject of another
talk.  Instead, we will try to argue that approximate moduli
spaces with less supersymmetry may not make sense.

\section{States with Less Supersymmetry}

At the classical level, we know of many approximate moduli spaces in
string
theory where supersymmetry is restored, if at all, only in limits
where there are an infinite number of gravitinos.  Examples
without supersymmetry include the ten-dimensional non-supersymmetric
theories
and toroidal
compactifications of these (no
supersymmetry restoration anywhere in the moduli space)
and compactifications (e.g. \cite{rohm}) where supersymmetry
is only restored as one takes $V\rightarrow \infty$.

There is some evidence that many of these theories are badly
behaved:
\begin{itemize}
\item  Most of these models have tachyons in some region of the
moduli space.  Thinking naively, this means that the energy
(cosmological constant) is unbounded below (even if the tachyon
potential has a minimum, the cosmological constant, at this point,
will be of order $-{1 \over g^2}$, and thus goes to $-\infty$ as
the coupling tends to zero).
\item
Many of these vacua are unstable, and undergo catastrophic decay
\cite{fabingerhorava}.  These decays are similar to the decay of
the Kaluza-Klein vacuum discussed long ago by
Witten\cite{wittenkkdecay}, in which a
rip appears in spacetime, which grows at the speed of light.
\item
Banks has argued, from considerations of black holes and
holography, that states without supersymmetry at least in some
asymptotic sense, may not be consistent\cite{littlelambda}.
\end{itemize}
These observations suggest that these theories might not make sense.
The problems seem generic.
But with our current state of knowledge, they are not decisive.  For
example, we could imagine that
stabilization occurs in a regime far from regions with tachyons, and
that
while the state is unstable, the lifetime is very long.  More generally,

in a
theory of gravity, as Susskind\cite{susskind} and
Banks\cite{banksisolated}
have stressed, it is not so easy to
decide when different states are part of the same theory.

\subsection{Speculations on Other Possible Problems for N=0:
Non-Perturbative Anomalies?}

We know that non-perturbative anomalies can render field theories
inconsistent.  It is natural to ask whether there might be such
anomalies in string theories, which might render some apparently
sensible states meaningless.  Such a possibility, indeed, was one
of the motivations for Witten's work which lead to the explosion
of interest in duality\cite{wittenusc}.

However, early on, Witten\cite{topologicaltools}
proved that there are no global anomalies in closed string theory, at
least
in the field theory limit under
rather weak conditions.  On the other hand, it is known that
non-perturbative anomalies
can lead to inconsistencies in open string theories; these problems are
in
some sense ``dual" to violations of modular invariance in closed
string theories\cite{famousauthors}.\footnote{Recently, deeper insight
into
these anomalies has been obtained by Uranga\cite{urganga}.}

Here we ask: are there additional consistency conditions in closed
string
theories?  Witten's result means that we need to look at states which
are
far from any smooth limit.
Below we briefly describe
a search for anomalies in discrete symmetries.

\subsection{Are All Discrete Symmetries Gauge Symmetries?}

It is widely believed that discrete symmetries in string theories are
gauged, so
anomalies would signal an inconsistency.
Examples include:
\begin{itemize}
\item  Discrete symmetries of toroidal compactifications:  these are
remnants of higher dimensional gauge symmetries.
\item  $E_8 \leftrightarrow E_8$ symmetry of the heterotic string:
This is a subgroup of a continuous group,
unbroken on a subspace of moduli space.
\item  T-duality of the heterotic string is, in many instances, a
subgroup of a continuous gauge symmetry\cite{tduality}.
\item  CP:  In many compactifications, one can see
that CP is a combination of a higher dimensional Lorentz
transformation and an ordinary gauge
transformation\cite{cpdiscrete}.
\end{itemize}

In other cases, it is less clear whether discrete symmetries are gauge
symmetries.
For example, some discrete symmetries of asymmetric orbifold theories,
and $S$-dualities are not easily recognized as gauge symmetries.
A general strategy to address this question might be the
following:
construct cosmic string solutions (and
analogs in higher dimensions), and ask what happens if one
moves particles (or more generally suitable p-branes) around them.
If they pick up suitable $Z_N$ phases (in the case of $Z_N$
discrete symmetries), then the symmetries are gauged.

One construction of such cosmic strings is due to Bagger, Callan
and Harvey\cite{bch}.  Their construction generalizes trivially to
any string vacuum described by a conformal field theory with
discrete symmetry $Z_N, N=2,3,4,6$.  The idea is to
compactify the $x_2,x_3$ directions on a {\it very} large torus with
$Z_N$
symmetry, modding out by the product of the internal $Z_N$ and the
$Z_N$ of the torus.   The fixed points of the group action are cosmic
strings.  The internal $Z_N$ still acts non-trivially on the states
of the theory.  It is easy to see that particles with $Z_N$ charge
indeed pick up a $Z_N$ phase as one moves about these fixed
points\cite{dgt}.  The main difficulty with this construction is
the fact that these strings have a very large tension and
correspondingly a large deficit angle.  As a result, it is not
totally obvious that these cosmic strings can be thought of as
topological objects sitting in the original vacuum.  Still, this
construction is highly suggestive.  An alternative construction,
suggested
by Banks, is under study.

\subsection{Searches for Anomalies}

The simplest way to search for anomalies in such symmetries is to
examine instanton effects in the low energy effective field
theory.  One asks whether the instanton determinant
violates the discrete symmetry.  One must be careful, however,
because it is often possible to cancel anomalies by a
Green-Schwarz mechanism, in which one assigns a transformation law
under the discrete symmetry to a modulus.  Such a search was
carried out in the past\cite{macintire} examining symmetric
orbifolds with supersymmetry.  A broader search is currently in
progress, including symmetric orbifolds without supersymmetry and
asymmetric orbifolds, both with and without
supersymmetry\cite{dgt}.
Currently, we have several examples of anomalies in both supersymmetric
and non-supersymmetric,
asymmetric orbifolds.  It seems quite possible that these
indicate the existence of new non-perturbative consistency
conditions in string theory.  If so, it would be important to
understand the stringy statement (analogous to modular invariance)
of the difficulty.   It is a bit disappointing that this test
doesn't seem to distinguish in an obvious way
between supersymmetric and non-supersymmetric
theories.
energy

\section{What does Supersymmetry Have to Say about $\sin(2\beta)$}

Since this talk was presented, there have been important
experimental developments.  In the original talk, I argued that
there were two generic predictions for CP violation in the $B$
system in supersymmetric theories:  ${\cal O}(10^{-2})$
or the standard model result (give or take
about $20\%$).  The reasoning was quite simple\cite{dns}:
\begin{itemize}
\item  If the $A$ parameter of order one, one needs CP violating phases
of
order $10^{-2}$ to understand $d_n$.  In models with supersymmetry
broken at an intermediate scale, one expects that the $A$
parameter is large, in general, so a natural explanation of the
smallness of $d_n$ would be that CP-violating phases are small.
In theories of lower scale breaking, particularly gauge mediation,
it is quite common for the $A$ parameter to be very small, so it
is not difficult to understand $d_n$.
\item  CP must be spontaneously broken in string theory.  Small
spontaneous breaking -- approximate CP -- can occur naturally
in such theories (simple models of this phenomenon are presented in
\cite{dns}).
This is consistent with the high scale breaking picture in the
previous item.  Nir has dubbed this situation ``approximate CP."
\item
In theories in which supersymmetry is broken at an intermediate
scale (supersymmetry breaking at low energies suppressed by
$1 \over M_p$), one doesn't expect excessive degeneracy of squarks
and
sleptons.  A review of models for understanding suppression of
flavor violation in supersymmetric theories was
performed in \cite{dns}.  Typically degeneracies of squarks
are at best of order one loop factors.  As a result, the real part
of $K-\bar K$ mixing is nearly saturated by supersymmetric
contributions in such models, and small phases can
(should?) be responsible for the observed value of $\epsilon$.
This is the case for flavor symmetries, ``dilaton
dominance," and ``gaugino dominance"\cite{gauginodomination}, and also
for ``anomaly
mediation"\cite{anomalymediation} and
``gauginomediation"\cite{gauginomediation},
which are often claimed to
yield much higher degrees of degeneracy (indeed, as explained in
\cite{adg}, there is, generically, no degeneracy).
\item
If CP is approximate, one can worry that
$\epsilon^{\prime}\over \epsilon$ will be too small.
It was argued in \cite{dns}, however, that
$\epsilon^{\prime}\over \epsilon$  can be of order the observed size if
some of the ``chirality breaking" squark mass terms are large.
This is
perhaps the most troubling aspect of this proposal, however.  The
terms mixing $\tilde s$ and $\tilde  d$ must be much
 larger than $m_s \theta_c$.  This is a natural phenomenon in theories
of
flavor, if some of the susy-breaking fields carry flavor quantum
numbers.  Still, it would be surprising.
\item
In theories with low scale breaking (e.g. gauge mediation),
$A$ terms are typically very small, and the degree of squark
degeneracy is very large.  So supersymmetry contributions to $\epsilon$
are small, and most of the contribution must come from the Standard
Model
processes.  There is no problem with understanding $d_n$, and
one expects results for $\sin(2\beta)$ close
to the predictions of the standard model.
\end{itemize}

\subsection{ The Experimental Situation}

Since this talk was presented, both Babar\cite{babar} and
Belle\cite{belle} have announced
results with greatly improved statistics for $\sin(2 \beta)$:
\beq
{\rm Babar}:
\sin(2 \beta) = 0.59 \pm 0.14.
\eeq
\beq
{\rm Belle}:
\sin(2 \beta) =0.99 \pm 0.14 \pm 0.06.
\eeq
This is to be compared to the standard model
expectation\cite{standarda}:
\beq
\sin(2 \beta) = 0.75 \pm 0.2.
\eeq

So it appears that,
if supersymmetry is present, the low scale option is more
promising.

\section{The Problem of Vacuum Selection}

One of the remarkable features of string theory is that,
classically, $N=1$ supersymmetry emerges so easily.  The
earliest examples of this kind were Calabi-Yau compactifications
of the Type I and heterotic string theories, but the bestiary
of such states quickly grew:  orbifolds, fermionic models, and
more intricate possibilities more recently (e.g. F-theory). In the
examples we know, there is always a moduli space of vacua, classically.
Many desirable features appear, such as chirality, generations,
intricate discrete symmetries, candidate hidden sectors, and the
like.

There are difficulties, however, both practically and conceptually,
associated with the fact that
there are so many states.  These states are labeled by both discrete and

continuous labels.  So there is a vast set of states to survey,
if one simply wants to look for one with a sensible phenomenology.
And if one wants to determine some dynamics which fixes among
these choices, almost inevitably the minimum of any potential for
the moduli must lie at strong coupling or zero coupling.

These issues are illustrated by the Horava-Witten limit of string
theory, the first of the ``brane world" pictures.  In strong
coupling,
the question is what fixes $R_{11}$ and $R$ ($S$ and $T$).
Phenomenologically, one wants, in order to understand unification
and the value of the Planck scale\cite{wittency},
\beq
R_{11} \sim 10-30 =  ~~~~~~ R^6 \sim 60
\eeq

This is an appealing picture.  But it leads to a puzzle.  For
large $R_{11}$, the theory is approximately five dimensional,
with $N=1$ supersymmetry in {\it five dimensions}.  But this is
enough to forbid a potential for $R_{11}$.  So one expects
stabilization, if at all, for $R,R_{11} \sim 1$.

There are some models for how this might come about:  Kahler
stabilization
and the
racetrack (reviewed in \cite{dinereview}), and discrete
fluxes\cite{fluxes}.  No picture is
yet totally satisfying.

\section{Coupling Unification in String Theory}

One of the triumphs of weakly coupled heterotic string theory is that
whether or
not one has a conventional grand unified structure at some energy
scale, unification
of couplings is {\it generic}.   But if the string coupling is not
weak, why should the couplings be unified?

The Horava-Witten picture illustrates that things
are not simple.  In general, the couplings do not unify.
Several moduli ($S$ and $T$) with large expectation
values couple to gauge fields, and the
$T$ couplings are not universal.
\beq
f_a \approx S + c_a T
\eeq
where $c_a$ is a numerical coefficient.
There are still large classes of theories for which
unification occurs.  In both the Kahler stabilization
and racetrack pictures, one can argue for unification.
But it is fair to say that we do not currently understand
in what sense coupling unification is generic.

\section{What Physics Might Choose the True Vacuum}

We have argued that there are many good vacua of string theory
with totally unacceptable properties for the description of
nature:  more than four dimensions, unbroken supersymmetries,
etc.  There are even vacua with $N=1$ supersymmetry with
unacceptable gauge groups\cite{banksdinemoduli}.  What has not been
established
is that there are states closely resembling the standard model.

Suppose there are?  Suppose Kahler stabilization, the racetrack,
or something else, produces a small number of stable vacua with
approximate $N=1$ supersymmetry, vanishing cosmological constant,
and with certain good
properties (coupling unification, proton stability,...).  Why
do we find ourselves in one of these?
It seems to me that if this is the picture, the answer is likely
to be at least partially anthropic.  One can, in fact, argue
that many of the vacua with too much supersymmetry will come
to a bad end:  structure will not form, for example.  Perhaps
one can make even stronger negative statements.  Inflation is
probably necessary in an anthropic view, but it is unlikely to be
generic\cite{banksinflation}.
So perhaps
there are only a small number with anything even remotely
resembling the possibility of life.

We might imagine that the universe, in its history, samples all
of these states, but that only this small subset is cosmologically
interesting (if we are lucky, perhaps many of the unappealing
ones never grow large, for example).  So there might be only
a small number of states which develop size and structure,
and we live in this one because it is the only place we can
live.  The racetrack model, for example, is suggestive of this:
stabilization only occurs for these theories for some finely-tuned
discrete parameters.  So acceptable states might not be generic.

While I won't explore this possibility further here, I think that many
of
the questions about the cosmology of the unappealing vacua
may be accessible now, and that this is an avenue worthy of
further exploration.  It is interesting to note that in the
proposal of Banks mentioned earlier, vacua with more than four
supersymmetries are unacceptable, because they possess infinite
numbers of states.

\section{The Brane World From the Perspective of String
Theory}

Prior to 1995, there were several proposals of very large
(TeV scale or larger) extra dimensions\cite{earlylargedimensions}.
But it was hard to make
sense of them in string theory, since the couplings were typically
{\it gigantic}.  With the understanding of duality, and
particularly of the role of various types of branes in string
theory, it became feasible to meaningfully discuss the possibility
of large or warped extra dimensions\cite{largedimensions,rs}.The
Horava-Witten
model was the first example.

>From our discussion of the Horava-Witten theory, some of the
issues facing a brane world picture are clear:
\begin{itemize}
\item  Stabilization of the moduli -- in general, one expects
stabilization when the moduli are of order one.  The most
plausible suggestion to understand larger radii is that the
large dimensions are effectively two dimensional, with
supersymmetry in the bulk of spacetime.   Then the potential for
$R$ is a function of $\ln(R)$, so the problem is just to explain
why this logarithm is $20$ or so.  In this picture, low energy
supersymmetry, as conventionally understood, is not an outcome.
\item
Suppression of dangerous processes:  in string theory, there are
no global continuous symmetries, and the effective theory
typically contains a huge array of higher dimension/derivative
operators, even at tree level.  Rare processes, precision electroweak
physics,
etc. must be understood in light of this.  Proposals include symmetry
breaking
on distant branes, discrete symmetries, and
others.
\item
If these problems are solved (by nature, if not by us), there is a rich
array of possible
new phenomena (see Dimopoulos's talk at this meeting).
The fundamental scale might be accessible to colliders.
\end{itemize}

\section{Conclusions}

While I can't claim to have offered any complete
answers in this talk, I have at least tried to bring into
focus some generic issues, which, if resolved, might lead to string
predictions:
\begin{itemize}
\item  What is the role of approximate moduli in string theory
\item  Can we rule out non-susy vacua?
\item  What lessons can we take from cosmology, and what
predictions can we make?
\item  Can we phrase analogous generic questions for large
dimensions, Randall-Sundrum?
\end{itemize}
Lurking in the background are many issues not addressed,
especially the {\it Cosmological Constant}.

\noindent
{\bf Acknowledgements:}

\noindent

This work supported in part by a grant from the U.S.
Department of Energy.  I wish to thank my collaborators on the
projects discussed here, particularly Alexey Anisimov, Michael
Graesser, Josh Gray, Yossi Nir and Yael Shadmi.
I would also like to thank Tom Banks for
many discussions of issues raised here.


\end{document}